# CONSIDERATIONS ON WAKEFIELD EFFECTS IN A VUV FELO DRIVEN BY A SUPERCONDUCTING TESLA-TYPE LINAC*

Alex H. Lumpkin†, Fermi National Accelerator Laboratory, Batavia, IL 60510 USA
Henry P. Freund[1], University of New Mexico, Albuquerque, NM 87131 USA
Peter Van der Slot, University of Twente, Enschede, the Netherlands
[1]also at University of Maryland, College Park, MD 20742 USA

*Abstract*

The effects on beam dynamics from long-range and short-range wakefields from TESLA-type cavities are considered in regard to a proposed FEL oscillator (FELO) operating at 120 nm. This would be driven by the Fermilab Accelerator Science and Technology (FAST) linac at 300 MeV with a 3-MHz micropulse repetition rate. Our wakefield studies showed measurable effects on submacropulse centroid stability and on submicropulse head-tail kicks that can lead to emittance degradation. In the case of the former, we use MINERVA/OPC to simulate the ~100-μm centroid slew effects on the saturated output power levels of the FELO.

## INTRODUCTION

The electron-beam properties needed for successful implementation of a free-electron-laser oscillator (FELO) on a superconducting TESLA-type linac at the Fermilab Accelerator Science and Technology (FAST) facility include the intrinsic normalized emittance and the submacropulse centroid stability [1]. We have demonstrated that short-range wakefields (SRWs) and long-range wakefields (LRWs) including higher-order modes (HOMs) are generated for off-axis beams in the two, 9-cell capture cavities and eight, 9-cell cavities of a cryomodule in the FAST linac [2-4]. The resulting degradation of the emittance and centroid stability would impact the FELO performance. At 300 MeV and with the 4.5-m long, 5-cm period undulator, the saturation of a vacuum ultraviolet (VUV) FELO operating at 120 nm has previously been simulated with GINGER and MEDUSA-OPC using the non-degraded beam parameters [1]. The measured electron-beam dynamics due to the SRWs (submicropulse, 100-micron head-tail kicks) and HOMs (submacropulse centroid slew of up to 100s of microns) will be presented [2-4]. These are mitigated by steering on axis as guided by the minimization of the HOM signals and beam dynamics effects. Simulations using MINERVA/OPC [5-7] of the effects of a submacropulse centroid slew on FELO performance will also be reported for the first time.

## EXPERIMENTAL ASPECTS

The Fermilab Accelerator Science and Technology (FAST) facility includes the superconducting TESLA-type linac [8] which could generate the driving beam at 300 MeV. This would enable an FELO operating at 120 nm. A schematic of the linac is shown in Fig. 1, and the photocathode rf gun. the two TESLA capture cavities operating at about 20 MV/m gradient, the first chicane for bunch compression, the cryomodule (with eight, 9-cell cavities) and the high energy transport line to the location of the FEL experiment are shown. The nominal electron beam properties are given in Table 1. The FELO experiments would be based on the U5.0 planar undulator [9,10] in hand with a 5.0-cm period, tunable magnetic gap, and 4.55-m length as summarized in Table 2. A schematic of the resonator cavity positioned in the high energy transport end of the beamline is shown in Fig. 2. In practice, the second 4-magnet chicane could provide e-beam bunch compression as well as access for the upstream mirror and be placed closer to the D600 dipole with the undulator downstream of this dipole. There is a second dipole, D603 (not shown), that would direct the electron beam off the optical axis and to the high-energy absorber. The 50-m optical cavity round-trip time matches the 3-MHz micropulse repetition rate. The downstream mirror was assumed to have about 80% reflectance at 120 nm, and we would use a 1-mm radius hole outcoupling.

Table 1: FAST Electron Beam Parameters.

| Parameter | Unit | Value |
| --- | --- | --- |
| Charge | nC | 0.5-1.0 |
| Emittance norm. | mm mrad | 2-5 |
| Gun energy | MeV | 4.5 |
| Beam energy | MeV | 300 |

Table 2: Summary of the U5.0 Undulator parameters [7].

| Parameter | Unit | Value |
| --- | --- | --- |
| Period | cm | 5.0 |
| K value | ….. | 0.45-3.9 |
| Length | m | 4.55 |
| Tunable gap | cm | 1.4-2.17 |
| Maximum field at 1.4 cm | T | 0.89 |

Past experiments on wakefield effects in the cavities which are correlated with their HOM signal strength motivated our instrumenting detectors on the HOM couplers for all 10 cavities. The HOM detectors are based on passband filtering on the first two HOM dipole bands from 1.6 -1.9 GHz and Schottky detectors as described elsewhere [2,4].

* Work supported by the U.S. Department of Energy, Office of Science, under Contract No. DE-AC02-06CH11359.
† lumpkin@fnal.gov

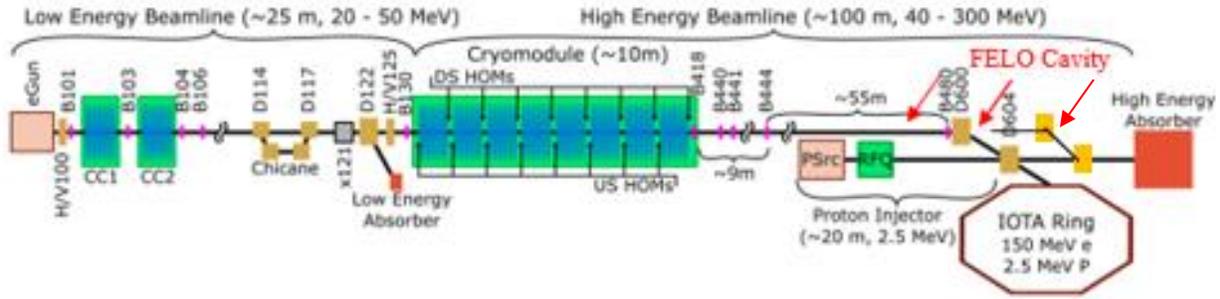

Figure 1: A schematic of the FAST linac showing the gun, capture cavities CC1 and CC2, the chicane, the cryomodule with US and DS HOM coupler locations, and the high energy transport to the FELO (see Fig. 2)

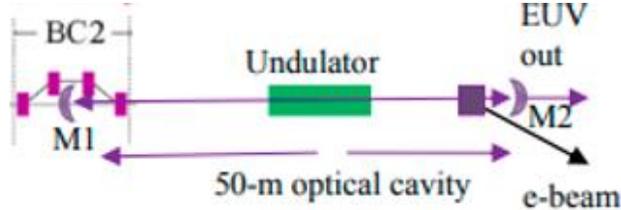

Figure 2: Schematic of a potential configuration for the FELO in the high energy transport area. A chicane would be added to provide access for the upstream mirror of the resonator.

## BEAM DYNAMICS FROM WAKEFIELDS

### Long-range Wakefields and HOMs

In the oscillator configuration, the submacropulse centroid slewing effects due to near-resonances of an HOM with a beam harmonic should be mitigated by careful, on-axis steering through the cavities. Examples of our observations after the cryomodule at 100 MeV energy are shown in Fig. 3. In the 50-b pulse train over 16.6 µs we see noticeable submacropulse centroid slewing in the bunch-by-bunch rf beam-position-monitor (BPM) data, particularly in B441 data with 300-µm total slew for -1 A (-2 mrad) steering [4]. We see the direction of the slew was correlated with the beam steering of the corrector V125 located 4 m before the CM. The corrector strength is 2 mrad/A. We also observed correlated centroid slew at location B480 located 64 m downstream of the cold BPM and just before the D600 dipole. The charge per micropulse was only 125 pC, and the kick effect goes directly with charge and inversely with beam energy. The higher charge in the FEL experiments and the 3x higher beam energy would roughly cancel these scaling effects so we ran our simulations initially at 100 µm total in 100 passes as reported in the next section.

### Short-range Wakefields

The short-range wakefields also could be a concern so proper on-axis steering of the beam through the 10 cavities is warranted to avoid head-tail centroid shifts within each micropulse. An example of the SRW kick from the capture cavity CC2 at about 35 MeV is shown in Fig. 4 where corrector V103 was used to steer the beam into CC2 [3,11]. The data were taken at the OTR screen at the down-stream location X121 as shown in Fig. 1. The magnitudes of the observed head-tail kicks versus charge and different angular steering up to ± 4 mrad are shown. At 1.5 nC of charge, head-tail kicks up to 300 µm were seen. The corresponding projected profiles from the y-t data increased up to 650 µm from 400 µm [11]. This would directly correspond to an emittance-dilution effect.

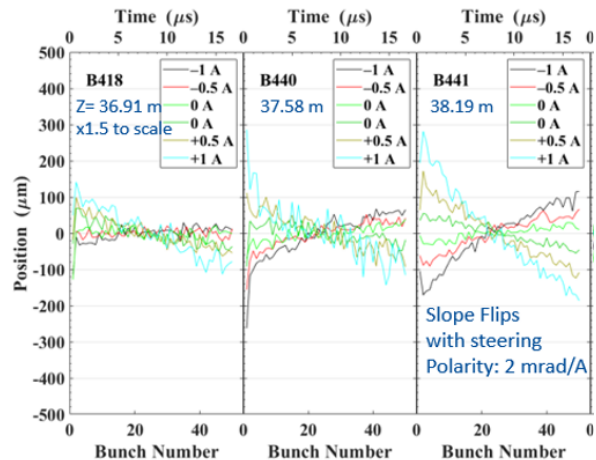

Figure 3: Submacropulse vertical centroid motion in the cold BPM and first two BPMs just after the CM due to HOMs excited by off-axis beam steering.

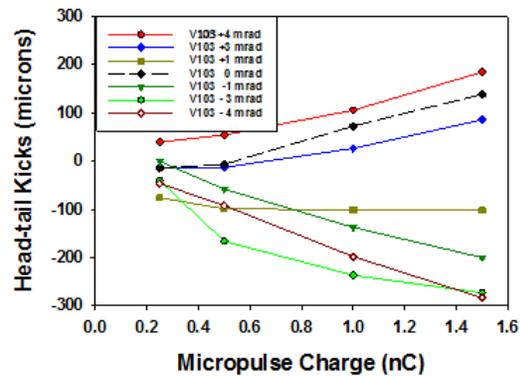

Figure 4: Submicropulse vertical centroid effect as the y-t head-tail kick for the streak camera images obtained at OTR screen X121 versus charge and V103 steering angle.

## MINERVA/OPC SIMULATIONS

The simulations were done in the steady-state regime with only a single temporal slice so there is no issue with the cavity detuning. The MINERVA simulations start with shot noise so that no initial seed is provided [5-7]. The basic parameters of the simulations are listed in Table 3 below for the FELO at 120 nm using the 5.0-cm period undulator.

Table 3: Parameters used in the simulation.

| Electron Beam | |
|---|---|
| Kinetic Energy | 301.25 MeV |
| Peak Current | 100 A |
| rms Energy Spread | 0.0005 |
| Normalized Emittance | 2.0 mm-mrad |
| Beam Size $x$ (rms) | 122 µm |
| Beam Size $y$ (rms) | 120 µm |
| Twiss $a_x$ | 1.0 |
| Twiss $a_y$ | 1.0 |
| Repetition Rate | 3.0 MHz |
| **Undulator** (flat pole face) | wiggle plane in $x$ |
| Period | 5.0 cm |
| Magnitude | 2.4614 kG |
| Length (1 period entry/exit taper) | 90 $\lambda_w$ |
| **Resonator & Optics** | concentric, hole out-coupling |
| Wavelength | 120 nm |
| Cavity Length | 50 m |
| Mirror Curvature | 25.32 m |
| Hole Radius (downstream mirror) | 1.0 mm |

In order to study the effect of a slew on the oscillator performance, we have added the possibility of including a slew in the beam centroid at the initialization of MINERVA. In order to engage this option, we specify (1) the slew direction, (2) the number of bunches (one bunch for each pass) over which the slew will be applied, and the maximum slew displacement over the number of bunches, and (4) the displacement (from the axis of symmetry) at the start of the simulation. Note that the default setting is that the initial displacement is zero. Therefore, if we specify that there will be 100 bunches and that the maximum displacement will be 100 µm, then the displacement will increase by 1 µm on each pass. If it is required that the simulation goes beyond 100 passes, then this increase in the slew will continue at this rate for each additional pass.

Figure 3 shows the out-coupled power vs pass assuming that there is no initial slew on the electron beam. A steady state is reached after about 60 passes at a power level of about 7.5 MW, while the power incident on the downstream mirror is about 154 MW for an out-coupling of about 5%. The single pass gain in the steady state is about 65%. Also shown in the inset are the case of no slew (blue) and 100 µm slew (red) plotted with a linear vertical axis. The output power sags to 5.5 MW at the end of 100 passes

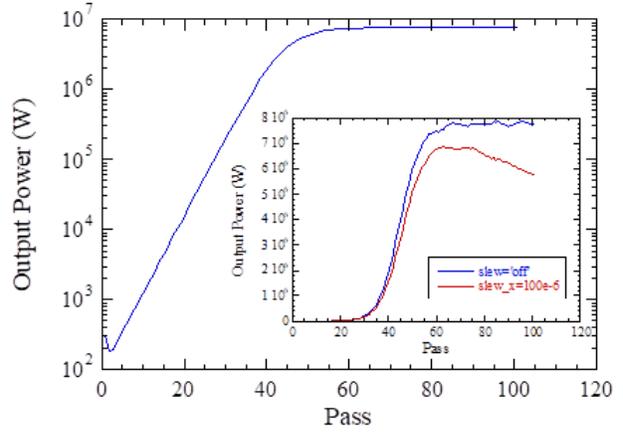

Figure 3: Plot of output power vs pass number at 120 nm with no slew with the inset showing on a linear vertical scale the power loss for a 100-µm slew (red curve).

under these conditions with initial displacement of zero in the x plane.

Figure 4 shows the effect of displacements in the x-direction which corresponds to the wiggler plane, and in the y-direction. It is clear from the figure that the effect of the slew is greater if it is in the wiggle plane. This is probably because there is no focusing in that direction in a planar undulator so the beam will remain off axis in that direction. While it might still excite radiation, the mode is probably also off axis and may not exit the resonator through the hole. It is also important to note that the degradation is only about 30% as the slew increases to 100 µm for a beam whose extent is about 120 µm. The fact that there is virtually no degradation in the performance when the slew is in the y-direction is probably due to the fact that there is focusing in that direction, and the beam will oscillate about the axis of symmetry. A mitigation effect occurs if the beam can be steered so it starts off axis at -1/2 the total slew in the x plane.

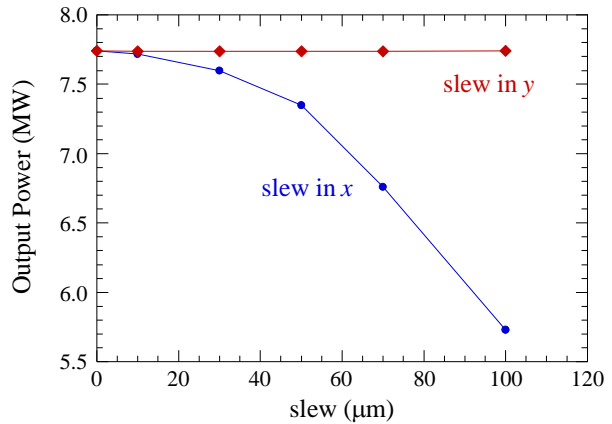

Figure 4: A comparison of the effects of slew in the x (or wiggle) plane (blue) and the y plane (red). The vertical focusing effects of the planar undulator mitigate the power loss for vertical slew.

## SUMMARY

In summary, we have considered for the first time the effects on the FELO performance at 120 nm should submacropulse centroid slews occur as may be driven by the HOMs in the TESLA-type cavities. Proper steering into the cavities to minimize the HOMs and at the undulator to make the slew symmetric in x should preserve FELO performance. These considerations would also apply to the recently proposed Tapering Enhanced Super-Radiant Stimulated Amplification (TESSA) based oscillator experiment at FAST at 515 nm [12].

## ACKNOWLEDGEMENTS

The first author acknowledges the support of C. Drennan and M. Lindgren of the Accelerator Division at FNAL.

## REFERENCES

[1] A.H. Lumpkin, H. P. Freund, M. Reinsch, "Feasibility of an XUV FEL Oscillator at ASTA", Proc. of FEL13, MOPSO51, New York, NY USA, 88-91, JACoW.org.

[2] A.H. Lumpkin *et al*., "Submacropulse electron-beam dynamics correlated with higher-order modes in Tesla-type superconducting rf cavities, Phys. Rev. Accel. and Beams **21**, p. 064401, 2018.

[3] A.H. Lumpkin, R.M. Thurman-Keup, D. Edstrom. J. Ruan, "Submicropulse electron-beam dynamics correlated with short-range wakefields in Tesla-type superconducting rf cavities", Phys. Rev. Accel. and Beams **23**, p. 054401, 2020.

[4] A.H. Lumpkin *et al*., "Submacropulse electron-beam dynamics correlated with higher-order modes in a Tesla-type Cryomodule", Phys. Rev. Accel. and Beams **25**, p. 064402, 2022.

[5] H.P. Freund, P.J.M. van der Slot, D.A.G. Grimminck, I.D. Setya, and P. Falgari, "Three-Dimensional, Time-Dependent Simulation of Free-Electron Lasers with Planar, Helical, and Elliptical Undulators", New J. Phys. **19**, 023020, 2017.

[6] H.P. Freund, P.J.M. van der Slot, and Yu. Shvyd'ko, "An X-Ray Regenerative Amplifier Free-Electron Laser Using Diamond Pinhole Mirrors", New J. Phys. **21**, 093028, 2019.

[7] P.J.M. van der Slot and H.P. Freund, "Three-Dimensional, Time-Dependent Ana lysis of High- and Low-Q Free-Electron Laser Oscillators," Appl. Sci. **11**, 4978, 2021.

[8] D. Broemmelsiek *et al*., " Record High-gradient SRF Beam Acceleration at Fermilab", New J. Phys. 20, 113018, 2018.

[9] E. Hoyer *et al*., "The U5.0 Undulator for the Advanced Light Source", Rev, Sci. Instrum. 63 (1), p. 359, 1992.

[10] P. Heimann *et al*., "Experimental characterization of ALS undulator radiation", Rev. Sci. Instrum. **66** (2), p. 1885, 1995.

[11] A.H. Lumpkin *et al*., "Direct Observations of Submicropulse Electron-Beam Effects from Short-range Wakefields in TESLA-Type Superconducting rf Cavities" Proc. of the International Beam Instrumentation Conference, IBIC20, TUPP17, JACoW.org.

[12] Alex Murokh and Pietro Musumeci, "FAST-GREENS: A High-Efficiency FEL Driven by a Superconducting rf Accelerator", presented at FEL2022, Trieste, Italy, August 2022, paper TUP33.